\documentclass[aps,prb,twocolumn,showpacs,amsmath,amssymb]{revtex4}
\usepackage{epsfig}
\usepackage{dcolumn}
\usepackage{bm}
\usepackage{amsmath}
\usepackage{amsfonts}
\usepackage{amssymb}
\usepackage{graphicx}%
\newcommand{\beq}{\begin{eqnarray}}
\newcommand{\eeq}{\end{eqnarray}}

\begin{document}

\title{Multifloquet to single electronic channel transition in the transport
properties of a resistive 1D driven disordered ring}
\author{Federico Foieri$^1$, Liliana Arrachea$^2$ and Mar\'{\i}a Jos\'e  S\'anchez$^3$.}
\affiliation{$^1$Departamento de F{\'i}sica ``J. J. Giambiagi" FCEyN, Universidad de Buenos 
Aires, 
Ciudad Universitaria Pab.I, (1428) Buenos Aires, Argentina,\\
 $^2$Departamento de F\'{\i}sica de la Materia Condensada and BIFI,
Universidad de Zaragoza, Pedro Cerbuna 12 (50009) Zaragoza,\\
$^3$Centro At\'omico Bariloche and Instituto Balseiro, Bustillo 9500 (8400), Bariloche, 
Argentina.}
\date\today
\begin{abstract}
We investigate the dc response of a 1D disordered ring coupled to a reservoir and driven by a 
magnetic flux with
a linear dependence on time. We identify two regimes: (i) A localized  or large length $L$ 
regime, 
characterized by a dc conductance, $g_{dc}$, whose probability distribution $P(g_{dc})$ is 
identical to the one exhibited by a 1D  wire of the same length $L$ and disorder strength placed 
in a Landauer  setup.  (ii) A ``multifloquet'' regime for small $L$ and weak  coupling to the 
reservoir,
which exhibits large currents and conductances that 
can be $g_{dc}> 1$, in spite of the fact that the ring  contains a single electronic 
transmission  
channel. The crossover length  between the multifloquet to the single channel transport regime, 
$L_{c}$, is  controlled  by  the coupling to the reservoir.
\end{abstract}
\maketitle
The  possibility to induce quantum transport on electronic devices driven by time dependent  
fields has 
concentrated in  recent years an impressive amount of  research. Some experimental examples 
along this 
direction are 
pumping phenomena \cite{pump1,pump2} and the development of ac generators in superconducting 
junctions\cite{pump3}. \\
\indent An early  proposal to generate a dc current and ordinary resistive behavior by recourse 
of time 
dependent (td) magnetic fields  was originally formulated by  B\"uttiker, 
Imry and Landauer \cite{bil,lanbu}.
They showed that a metallic loop  enclosing  a  magnetic flux varying linearly with 
time 
(a constant electromotive force), exhibits an  ohmic-like behavior in the current response when 
it is 
coupled to a dissipative environment like a particle reservoir.
By that time it was already recognized that the Landauer's formulation \cite{land}, provides a 
method to 
compute the steady state current of non interacting electrons in  mesoscopic samples  
connected   to 
electrodes at different chemical potentials. 
Therefore, a quantum wire with definite  characteristics like number of channels, length and 
disorder
strength, can be bended  in a ring shape in contact with a particle reservoir and  induce a 
current by 
the  
application of a td magnetic flux or it can be placed in a two terminal set up to establish a 
current 
through the application of a bias in the chemical potentials.   
A natural question that emerges is in which extent the obtained  dc current depends on the
way in which transport has been induced. 
Related discussions have been addressed in previous studies of the conductivity of a normal 
metal ring in the 
framework of Kubo formula with a phenomenological treatment of the effect of 
a dissipative environment \cite{trivedi, reulet}. 
In recent literature transport phenomena induced by  electric fields, 
without introducing different chemical potentials, have been discussed \cite{kohn} and 
expressions for the 
conductance equivalent to that of Landauer theory have been recovered within linear response in 
the ac electric 
field.\\
\indent The Landauer (dimensionless) conductance of a disordered wire $g_w$ displays large 
sample to sample 
fluctuations \cite{benrev}. Today the full distribution, $P(g_{w})$, is known for  quasi 1D 
and 1D 
wires, even at regimes of finite temperature and bias voltage \cite{fede1}.
Whether a similar universal characterization can emerge in the case of the
dc conductance induced by a td flux in a disordered ring is an open question that will be 
addressed 
in this work. \\ 
\indent We analyze a 1D disordered  ring of length $L$ threaded by a magnetic flux with a linear
 dependence on time  $\Phi(t)= \Phi t$ and coupled  to a reservoir. This introduces dissipation 
of energy 
 and  generates a dc current $J_{dc }$ along the ring \cite{lanbu}.  
Before giving further details, we anticipate here our  main findings. Two  regimes in the 
behavior of 
the dc current have been identified. The short length  transport regime is 
characterized by large dc currents
and (disordered average) $g_{dc}$ conductances  that, for  weak coupling to the reservoir,
  can be $g_{dc}= J_{dc } /\Phi > 1 $, in spite of the 
 fact that the ring  contains a single electronic transmission  channel. We name this regime 
``multifloquet", because 
 on top of the  discrete  electronic levels of the ring, additional resonances associated to the
harmonic dependence of the Hamiltonian with time contribute to the dc current. Within this 
regime, we 
find that
 $J_{dc } \sim L^{-\alpha_{c}}$, with $\alpha_{c}$ depending on the value of the  coupling to 
the reservoir $t_c$.
In the second regime, that we called single channel-Landauer regime, $g_{dc}$ and its associated  
probability distribution, $P(g_{dc})$, behave like those  of a 1D disorder  wire of the same 
length $L$ 
and disorder strength, placed between two reservoirs at a chemical potential difference $\Phi$. 
In this regime,
 $g_{dc}$ decays with $L$ following  an exponential law $g_{dc} \propto \exp(-L/l)$, 
with $l$ the localization length. Below, we 
estimate the crossover length, $L_c$, between both regimes. \\
\indent We investigate a setup given by the  Hamiltonian: \cite{liliring}
$H_{res}+ H_{cont}+ H_{ring}(t)$. The first term describes the reservoir,
which we assume to be a free electronic system in equilibrium with many degrees of freedom 
labeled by $k_r$,  with a well defined temperature $1/\beta$ and chemical potential $\mu$.
The Hamiltonian for the ring reads:
\begin{equation}\label{ham}
H_{ring}(t)=-t_{h} \sum_{l=1}^N \Big( e^{i \Phi t/N} c^{\dagger}_l c_{l+1}
+ e^{- i \Phi t/N} c^{\dagger}_{l+1} c_l  + \varepsilon_l c^{\dagger}_l c_{l} \Big),
\end{equation}
where $N$ is the number of sites along the ring, i.e. $L=N a$, being $a$ the lattice constant.
 The td phases account for the td flux and 
$\varepsilon_l$ denote the random on site  energies uniformly distributed in the disordered 
profile 
with zero mean
and width ${\cal W}$. The
hopping between nearest neighboring sites  is $t_{h}$ and  we adopt units in which the flux
 quantum is 
$\phi_{0}=1$, $\hbar=1$.
The  term $ H_{cont}= t_c\sum_{k_r}(c^{\dagger}_{k_r} c_1+ H.c.)$ describes the coupling 
between  
ring and reservoir in terms
of a hopping matrix element $t_c$ connecting the site $l=1$ on the ring with the reservoir.
We summarize the procedure introduced in Ref.\onlinecite{liliring} to compute the current 
employing the 
Keldysh non-equilibrium Green's functions formalism.
We first perform a gauge transformation 
$c_l \rightarrow c_l e^{-i l t \Phi/N } $, which transforms 
$H_{ring}(t) \rightarrow H_0 - {t_h} \Big( e^{- i \Phi t} c^{\dagger}_1 c_N + H. c. \Big)$,
being 
$H_0= -{t_h} \sum_{l=1}^{N-1} \Big( c^{\dagger}_l c_{l+1} + H. c.\Big) + 
\sum_{l=1}^{N}\overline{\varepsilon}_l
c^{\dagger}_l c_{l}  $, with $\overline{\varepsilon}_l=\varepsilon_l+ l\Phi /N$. 
Under this transformation  $H_{ring}(t)$ depends {\em harmonically} on time with a 
frequency  $\Phi$.
We consider a semi-infinite tight-binding chain
with hopping $t_h$ as the reservoir. 
The corresponding degrees of freedom can be integrated
out, which defines an equilibrium self-energy  determined  by the spectral function:
$\Gamma(\omega)$.
The retarded Green's function for the {\em transformed} Hamiltonian $H_{ring}(t)$ coupled to the
reservoir can be exactly calculated by solving Dyson's equation. In the present problem, 
the latter has the structure
of a  $1D$ disordered wire of length $L$ coupled to a reservoir at site $l=1$ and closed
by  an effective td hopping that couples electrons to effective
quanta of frequency $ \Phi$ at the bond 
$(1, N)$:
\begin{eqnarray}\label{dyson}
& & G^R_{l,l'}(t,\omega)\!\!=\!\! G^0_{l,l'}(\omega) -\!\!{t_h} e^{-i \Phi t}
G^R_{l,1}(t,\omega+ \Phi)G^0_{N,l'}(\omega)  \nonumber \\
& & -{t_h}  e^{i \Phi t}
G^R_{l,N}(t,\omega- \Phi)G^0_{1,l'}(\omega), 
\end{eqnarray}
being $G^0_{l,l'}(\omega)$  the
retarded Green's function of the equilibrium problem defined by 
$H_0+ H_{cont} + H_{res}$. \\ 
\indent New insight into the problem is obtained by introducing the Floquet representation
of Refs.\onlinecite{lilip}: $G^R_{l,l'}(t,\omega)=
\sum_{k=-\infty}^{\infty} e^{- i k \Phi t} {\cal G}_{l,l'}(k,\omega)$.
Using properties of the Fourier components 
of the retarded Green's function
${\cal G}_{l,l'}(k,\omega)$, the
dc-component of the current along  the ring can be exactly expressed as
$J_{dc}=\sum_{k=-\infty}^{\infty} J_{dc}(k)$, with:
\begin{eqnarray}
 J_{dc}(k) &=& 2 {t_h}  
\langle \int  \frac{d\omega}{2\pi}[f(\omega)-f(\omega- k\Phi)] \Gamma(\omega+\Phi/N ) 
\nonumber \\
 & & \!\!\!\!\!\!\!\!\!\!\!\!\!\!\!\!\!\!\!\!\!
\times \mbox{Im} \Big\{ {\cal G}_{l,1}(k,\omega+\Phi/N) 
[{\cal G}_{l+1,1}(k,\omega+\Phi/N)]^* \Big\} \rangle,
\label{jdck}
\end{eqnarray}
where $\langle \ldots \rangle $ denotes average over disorder. 
Due to the conservation of the charge, the above current does not depend on the
bond $( l, l+1 )$ considered for the calculations. The Fermi function
$f(\omega)= 1/(e^{\beta(\omega-\mu) }+1)$ depends on the chemical potential and
temperature of the reservoir. 
The superposition of $k$-components contributing to the dc current indicates that
not only electronic channels within a  small window of  width $\Phi$ around the Fermi
level contribute to the dc current, but  also `hot' electrons excited by the absorption of
$k$  quanta $\Phi$, as well as electrons deep in the Fermi sea.
Remarkably, for $k=1$, Eq.(\ref{jdck}) resembles the expression for the dc current   
induced in a 1D wire in a Landauer set up, {\em i.e.} with two reservoirs at  a chemical 
potential
difference $\Phi$, which can be exactly expressed as follows:
 \begin{eqnarray}
 J_{w} & = &  2 {t_h}
\langle \int \frac{d\omega}{2\pi} [f(\omega)-f(\omega - \Phi)] \Gamma(\omega ) \times
\nonumber \\
& & 
 \mbox{Im} \Big\{
G_{l,1}(\omega)
[G_{l+1,1}(\omega)]^* \Big\} \rangle,
\label{jdcw}
\end{eqnarray}
being $G_{l,l'}(\omega)$ the retarded Green's function for the Hamiltonian
$H_0$ of the disorder wire coupled at sites $l=1, N$ {\em to left and right reservoirs} 
respectively, with the same density of states and coupling $t_c$.\\
\noindent Below, we show that the solely condition to be fulfilled in order
to achieve single-channel conduction 
 in the ring (defined by the single contribution of $J_{dc}(k=1)$ to the total current)
is a low tunneling amplitude along the disordered
wire coupled to the reservoir. Formally, this condition implies that  the Green's function
behaves like  $G^0_{1,N}(\omega) \rightarrow 0$. In fact, under such an assumption, 
the set of Eqs. (\ref{dyson}) can be explicitly
solved and the result is:
\begin{equation}
G^R_{l,1}(t, \omega)= {\cal G}_{l,1}(0, \omega) + 
 {\cal G}_{l,1}(1, \omega) e^{- i \Phi t}
\end{equation}
being:
\begin{eqnarray}
\label{calg}
{\cal G}_{l,1}(0, \omega) &=& \frac{G^0_{l,1}(\omega)}
{1-{t_h}^2 G^0_{1,1}(\omega)G^0_{N,N}(\omega)},\nonumber \\
{\cal G}_{l,1}(1, \omega) &=& - {t_h} \frac{G^0_{l,N}(\omega)G^0_{1,1}(\omega) }
{1-{t_h}^2 G^0_{1,1}(\omega)G^0_{N,N}(\omega)} \;.
\end{eqnarray}
It is straightforward to prove that the function 
$G^{eff}_{l,l'}(\omega) \equiv {\cal G}_{l,l'}(0, \omega)+{\cal G}_{l,l'}(1, \omega)$ is the 
Green's function
of an equilibrium ring in contact with a reservoir at $l=1$, closed with a hopping
$t_h$ between $(1,N )$.
Thus for large enough $N$ and low transmission along the 
wire,  the latter  function becomes equivalent to  the Green's function
of the wire under consideration coupled to left and right reservoirs, i.e.
$G^{eff}_{l,l'}(\omega) \rightarrow G_{l,l'}(\omega)$.
Replacing $G_{l,l'}(\omega)$ by  $G^{eff}_{l,l'}(\omega)$ into Eq.(\ref{jdcw}) and taking into 
account that
solely  $\langle \mbox{Im}[{\cal G}_{l,1}(1, \omega){\cal G}_{l+1,1}(1, \omega)^*] \rangle$ 
contributes with a non vanishing value to the integral, we obtain Eq.(\ref{jdck})  with 
$k=1$ (up to terms ${\cal O}(\Phi/N)$ and boundary effects  which we suppose are vanishing small 
at large $N$ and low $\Phi$ values in which we shall concentrate in what follows).

\begin{figure}
\includegraphics[width=0.9\columnwidth,clip]{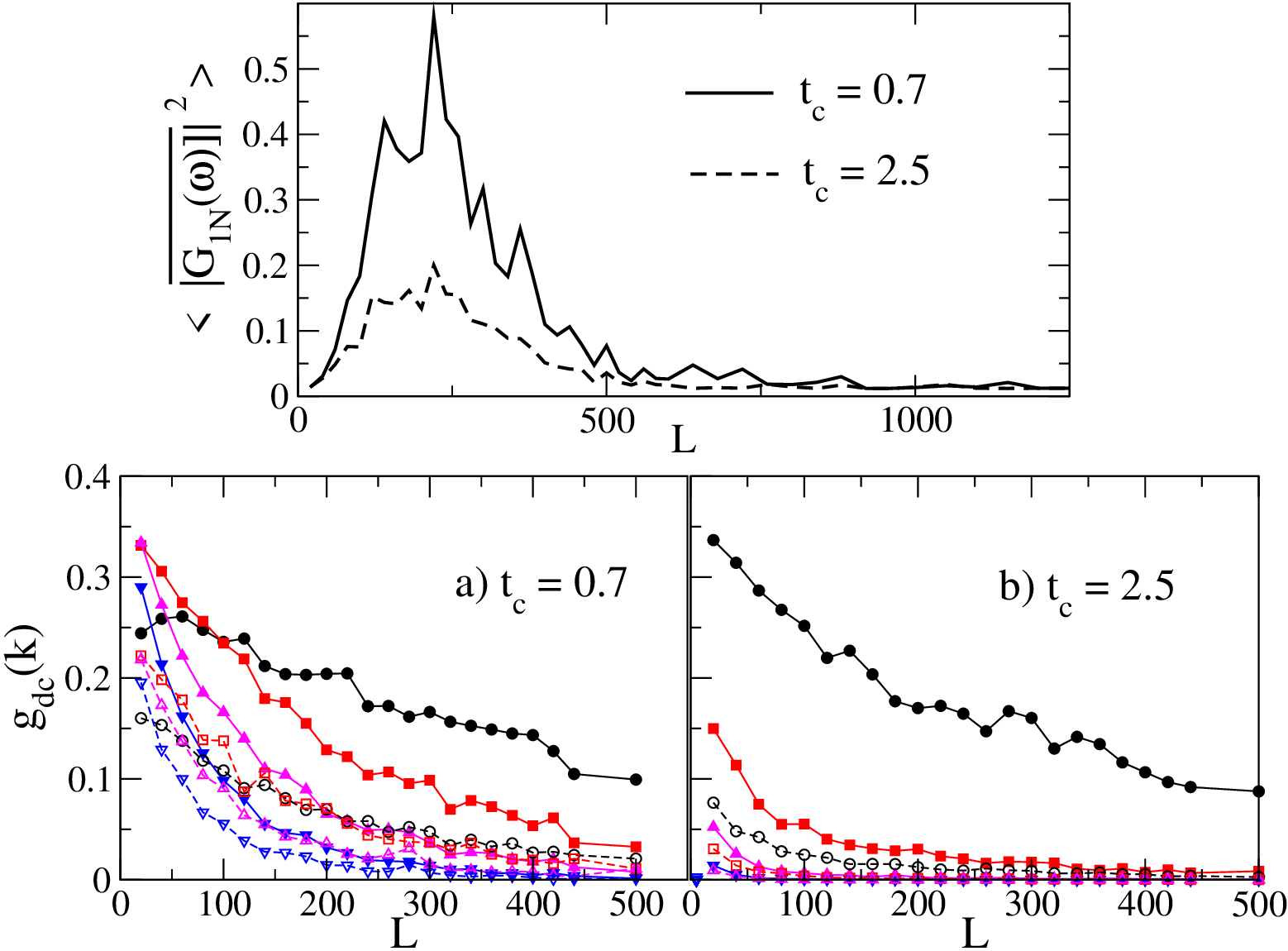}
\caption{\label{fig1} Top panel: 
 Disorder averaged of $  \langle \overline{ | G^0_{1,N}(\omega)|^2 } \rangle $ 
in a window of width
 $\delta \omega = \Phi=0.01 $ ($\overline {\lbrace...\rbrace}$ denotes frequency average)  as 
a function 
of the length $L$ for $t_c= 0.7$ (solid line) and $t_c= 2.5$ (dashed line).
Lower panel (Color on-line). Contributions  to the conductance
$g_{dc}(k)$  as a function of $L$  and for the same  $\Phi$, for
 $|k|= 1$ (black circles), $|k|=2$ (red squares), $|k|=3$ 
(magenta up triangles)  and $|k|=4$ ( blue down triangles). Solid (open) symbols
 correspond to $k > 0 (k < 0)$. 
a) $t_c= 0.7$ and  b) $t_c= 2.5$, for this coupling the single channel regime is reached 
for smaller values of $L$. The averages are over 5000 disorder realizations and for ${\cal W}=0.5$.}
\end{figure}

Therefore, for low transmission  
amplitude along the wire coupled to the reservoir, we obtain a single-channel dc current with
the ensuing conductance  equivalent to that obtained in a Landauer setup.
Physically, two ingredients provide the condition to achieve such a behavior. The first one
is  a strong coupling $t_c$ to the reservoir that enhances
for the electrons the probability of exiting to the reservoir instead of conducting 
along the wire. The second ingredient is  disorder, which favors localization along the wire.
We substantiate this on the basis of numerical results  that are displayed in
 Fig.\ref{fig1}.  We fixed ${\cal W} =0.5$ and focused on zero temperature.
In the upper panel it 
is plotted  $\overline{\langle | G^0_{1,N}(\omega)|^2 \rangle }$ averaged for $5000$ disorder realizations and 
in 
a window of width $\delta \omega = \Phi$ ($\overline { \lbrace...\rbrace}$ denotes frequency 
average)  
as a function of the length $L=Na$ for two values of coupling to the reservoir
 $t_c=0.7$ and $ t_c=2.5$. It is evident that, for  the 
 same  disorder, the  stronger  coupling favors  
that $ \langle \overline{| G^0_{1,N}(\omega)|^2 } \rangle \rightarrow 0$ at 
 smaller system size. This is concomitant with the behavior of the dc conductance, as it  is 
illustrated
in the lower panels of Fig.\ref{fig1}, where the relevant contributions
 $g_{dc}(k)\equiv J_{dc}(k)/\Phi$ (see Eq.(\ref{jdck})) are shown as 
functions of $L$ for  
 $t_c=0.7$ (panel a)) and  $t_c= 2.5$ (panel b)). In the latter case, 
besides the fact that all  the components $g_{dc}(k)$
 decrease monotonically with $L$, the dc response is mostly dominated by  $g_{dc}(k=1)$ for 
all $L$.  

\begin{figure}
\includegraphics[width=0.9\columnwidth,clip]{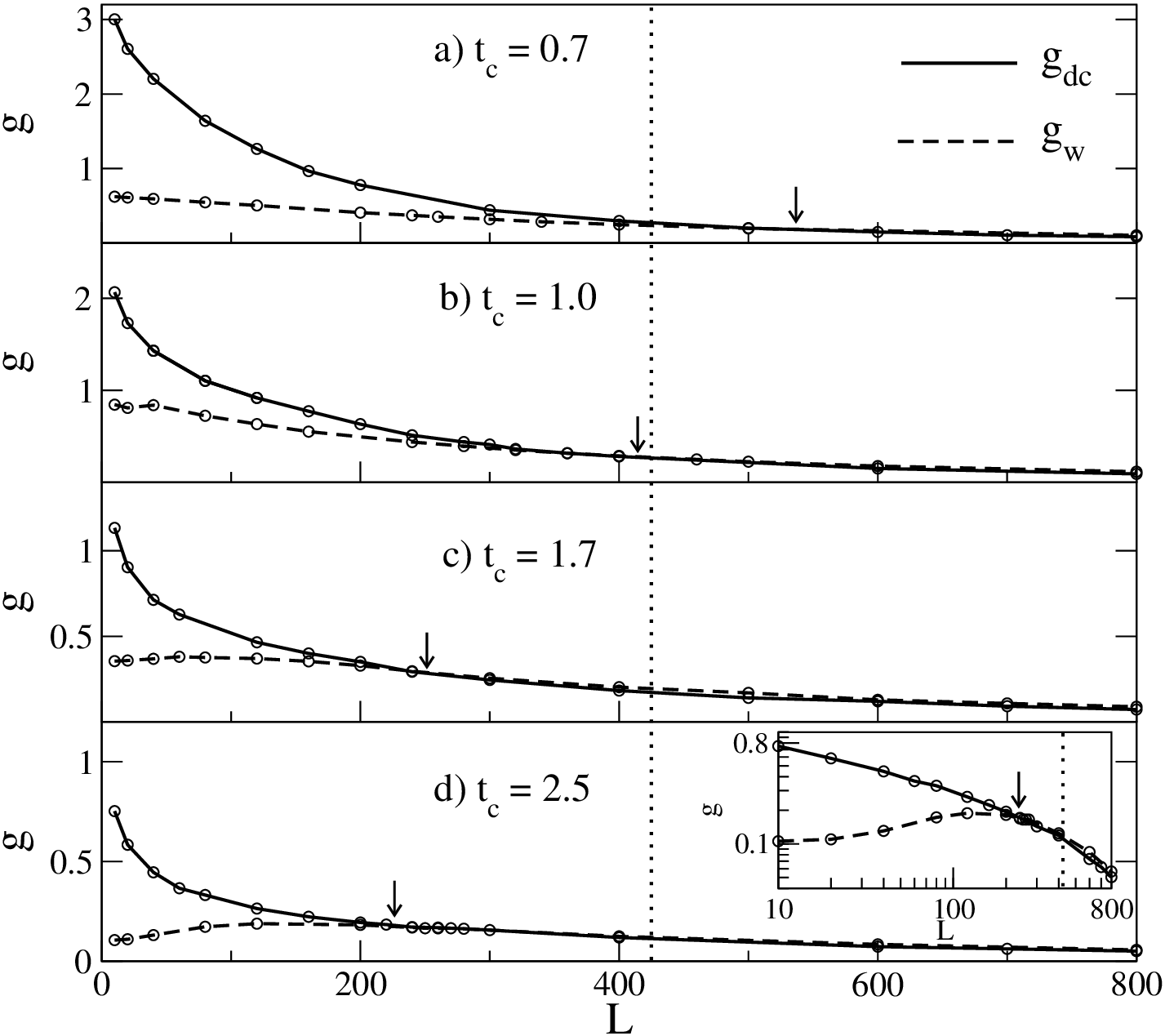}
\caption{\label{fig2} $g_{dc} $  
 for the disordered ring 
(solid line) and  $g_{w}$ for the $1D$ wire (dashed line) calculated with 
 Eq.(\ref{jdcw}) with chemical potential 
difference $\Phi$, as functions of $L$.
 The different panels correspond are a) $t_c= 0.7$, b) $t_c= 1$ c) $t_c= 1.7$ and c) $t_c= 2.5$.
 The localization length $l= 425$ for  $t_c = t_h = 1$ is shown as a vertical 
dotted 
line and the
crossover length $L_c$ is  indicated in each panel by  a vertical  arrow.
 Inset: Log-log plot of $g$ vs $L $  for $t_c= 2.5$. The transition in $g_{dc}$ from 
  the power-law  to the exponential decrease with $L$ is at the crossover  length $L_c \sim 230$.
  Other details are as in Fig. \ref{fig1}.}
\end{figure}

In Fig.\ref{fig2} we plot, as a function of $L$ and for different couplings $t_c$, the disorder averaged  conductances  $g_{dc}=J_{dc}/\Phi$ (solid line) for the ring coupled to the reservoir  and $g_{w}= J_{w} / \Phi$
 (dashed line) obtained from Eq.(\ref{jdcw}) for a 1D wire coupled to both reservoirs with $t_c$. 
 For each coupling, we  identify a crossover length $L_{c}$
at which the two conductances overlap for $L>L_{c}$. 
For $t_c=1$,
$L_c \sim l$, being $l$ the localization length  obtained for a wire with disorder
strength ${\cal W}$ (for the parameters under consideration $l=425$). 
For other couplings we found that $L_c/l$ is a 
monotonic decreasing 
function of $t_c$. We have verified that
for $L > L_c$ the regime is single-channel i.e. $ g_{dc} \sim g_{dc}(k=1)$ and that it follows
the  exponential behavior expected for a disorder wire, i.e. $g_{dc} \sim g_{w}\propto 
\exp{(-L/l)}$.
On the other hand, in the ``multifloquet'' regime for $L< L_c$, 
the two conductances differ dramatically and, in addition, can be  $g_{dc}>>1$. 
In this regime   we find a power law behavior $g_{dc} \propto L^{-\alpha_c} $, with the exponent
$\alpha_c$ slightly dependent  on the coupling to the reservoir.  
 This is illustrated in the inset of 
 Fig.\ref{fig2} panel d), where we show in solid line the log-log plot of $g_{dc}$ 
vs $L$ for $t_{c}= 2.5$.
Our estimates cast $\alpha_c=0.3 (0.4)$ for  $t_c=0.7 (2.5)$, respectively.

\begin{figure}
\includegraphics[width=0.9\columnwidth,clip]{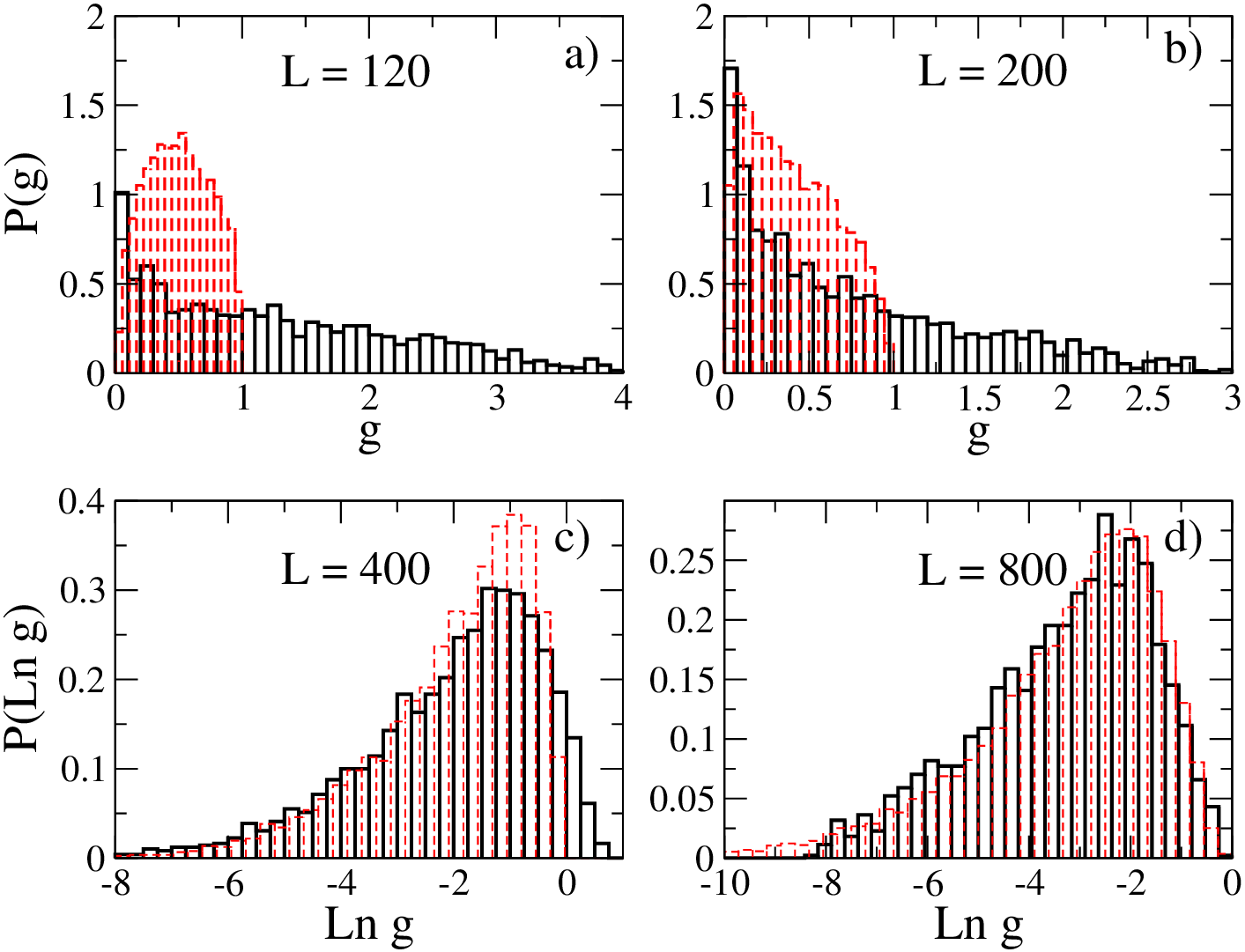}
\caption{\label{fig3} (Color on-line) Conductance distribution $P(g)$ for the disordered ring (black solid line) 
and  wire  (red dashed line) for $t_c= 0.7$
and a) L=120, b) L=200, c) L=400 and d) L=800. 
Other details are as in Fig. \ref{fig1}.}
\end{figure}

\begin{figure}
\includegraphics[width=0.9\columnwidth,clip]{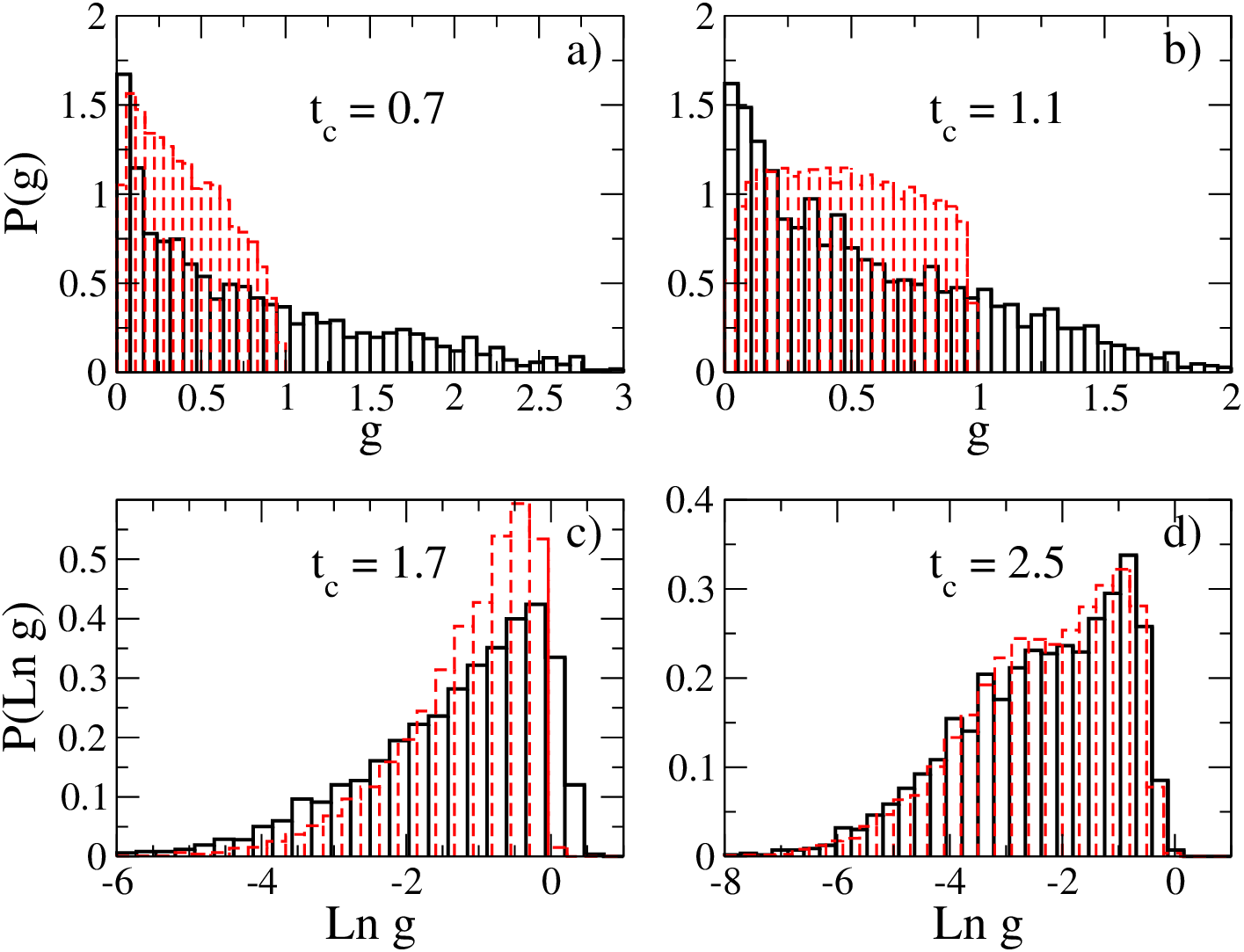}
\caption{\label{fig4} (Color on-line)
The same as Fig.\ref{fig3} for $L=200$ and
a) $t_c= 0.7$, b) $t_c= 1.1$, c) $t_c= 1.7$ and d) $t_c= 2.5$.}
\end{figure}

So far, we have focused in the mean values of the conductance. In what follows, we present 
further 
evidences of the multifloquet to single-channel transition,
 based on the analysis of the full distribution of the dc 
conductances. 
In Fig.\ref{fig3} we plot the distributions
$P(g_{dc})$ (solid line histogram) and $P(g_{w})$ (dashed line histogram) 
for a weak coupling to the reservoir, $t_c= 0.7$, for which we find $ L_{c} \sim 500 $. 
For lengths $L< L_{c}$, that is in the ``multifloquet'' regime, the  distribution $P(g_{dc})$
spreads over a wide range of conductance values, much larger than the maximum allowed 
for a single channel wire ($g^{max}_{w} = 1$). On the other hand, for  
$L=800 > L_{c}$ (panel c))
 both histograms correspond to the same  Log-normal distribution, characteristic of 1D 
disordered  wires in the localized regime \cite{benrev}. 
For completeness,  the distributions $P(g_{dc})$ (solid line 
histogram) 
and $P(g_{w})$ (dashed line histogram) at a fixed length $L=200$ and as a function of $t_c$ are 
shown 
in Fig.\ref{fig4}. For the strongest coupling considered, $t_c= 2.5$ (panel d)), the estimate is
$L_{c} \sim 230$ (see Fig. \ref{fig2}) and  in consistency, $P(g_{dc})$ for 
$L =200$ closely follows $P(g_w)$.

To conclude, we have identified two regimes in the dc response  of 
a 1D disorder ring threaded by a linear time-dependent magnetic flux
and coupled to a reservoir: (i) a ``multifloquet'' regime
that can give rise to  high dc currents, significantly larger than those expected in a Landauer 
setup with an equivalent bias, and  (ii) a single channel regime
with identical transport behavior as in a Landauer setup, not only in the average 
conductance but in the complete probability distribution as well. 
Our results have at least two important outcomes. The first one is that 
 the ``multifloquet'' regime can support  much higher
currents than those expected from the naive argument of assuming that only electrons at the Fermi level
contribute to the current. This regime should be conceptually equivalent to the ``ultra quantum'' regime defined in Ref. 
\cite{krav} and  could be realized, for example, in mesoscopic rings with a low level of disorder
and with a very low coupling to  the environment. Secondly, it would also take place in  rings driven by harmonically 
time-dependent fields, like  for example, a 
magnetic flux with an harmonic dependence on time, which corresponds to
 a Hamiltonian with  a similar structure as Eq.(\ref{ham})
\cite{note}. Thus, some conclusions of the present work, in particular the existence of 
the ``multifloquet'' regime, should also be valid
for Aharanov-Bohm mesoscopic rings with harmonically time-dependent fluxes, in 
which amplified values of the persistent currents, have been already measured \cite{bouchiat}. 
In addition some ab-initio calculations have been recently reported where the mechanism to 
introduce an electric field in the sample is exactly the one considered in the present work, namely, threading
the ring with a time-dependent flux \cite{dft}. It could be then interesting to check the present
predictions by alternative methods. 

We acknowledge support from CONICET, Fundaci\'on Antorchas PICT 0311609 and PICT 0313829, 
Argentina, as well as  FIS2006-08533-C03-02, the ``Ramon y Cajal'' program from MCEyC, 
grant DGA for Groups of Excelence of Spain and
the hospitality of Boston University (LA).

\begin{thebibliography}{99}
\bibitem{note} The frequency $\Omega_0$ of the ac flux  enters instead of  $\Phi/N$,
and additional terms involving higher harmonics of $\Omega_0$ would also contribute 
to (\ref{ham}).
\end{thebibliography}
\end{document}